\documentclass[%
 reprint,
 amsmath,amssymb,
 aps,
]{revtex4-2}

\usepackage{graphicx}
\usepackage{dcolumn}
\usepackage{bm}

\begin{document}


\title{Pathological limits in statistical mechanics}

\author{C.Y. Chen}
 \altaffiliation{Physics Department, Beihang University, PRC.\\ email: cychen@buaa.edu.cn  } 
%

\date{\today}

\begin{abstract}
This paper shows that some of the limit-like quantities currently used in statistical mechanics are  
ill-defined in the mathematical sense. Along the line, it is shown that significant progresses in non-equilibrium gas dynamics can be made  by redefining, reinterpreting, and reformulating  those quantities. 
\end{abstract}



\maketitle
\newcommand{\beq}{\begin{equation}}
\newcommand{\eeq}{\end{equation}}
\newcommand{\ba}{\begin{eqnarray}}
\newcommand{\ea}{\end{eqnarray}}
\newcommand{\pt}{\partial}
\newcommand{\lsigma}{\sigma^l}
\newcommand{\act}{\lfloor}
\newcommand{\db  }{\delta\beta}
\newcommand{\de  }{\delta}
\newcommand{\br  }{{\bf r}}
\newcommand{\bv  }{{\bf v}}

\section{\label{sec:level1}Introduction}

In mathematics the term ``pathological object'', or ``monster'',  refers to something whose behavior is  unexpectedly bad, counter-intuitive, and inexplicable in terms of  supposedly relevant theories. As a matter of fact, many of the so-called ``counterexamples'' in the books titled ``Counterexamples in analysis", ``Counterexamples in topology" and ``Counterexamples in probability"\cite{B1,B2,B3} can be deemed as typical pathological objects.

 Part of the reason why mathematics, known as the queen of all sciences, has in itself so many troublesome objects is that basic concepts in analysis (calculus), such as infinitely small, infinitely large, limit, and continuity, are simple but complex. When described in the daily language, they seem  elementary and plain; however, when defined as rigorous and widely applicable concepts, they 
 must employ a certain degree of abstraction and indirectness. If they are used to build more sophisticated structures in mathematics or other sciences, the abstraction and indirectness involved in them may become obscure sources of mishandling or misunderstanding. Professional mathematicians are fully aware of this type of risk. When they come up with or encounter a new theory in mathematics, they often spend a lot of time and effort, much more than expected, finding out and clarifying every possible ambiguity related to the theory. 

Not surprisingly, non-mathematicians pay much 
less attention to this aspect of mathematics.
In the physics community, there is a saying like this: mathematics is a tool  unreasonably effective in natural sciences\cite{wigner}. Although the saying contains a lot of wisdom, inspired and continues to inspire many remarkable developments in physics, the other side of the issue cannot and should not be ignored completely.
To promote a balanced awareness of the issue,  
this paper presents a case study showing that pathological objects in mathematics and 
half-baked ideas in physics have natural tendency to find each other and form seemingly 
plausible misconceptions (reflecting what  Murphy's law tries to express: if anything can go wrong it will).

The concrete objective of this paper is to unveil that some of the multi-variable limits currently used in non-equilibrium statistical mechanics are ill-defined in mathematics and ill-behaved in the physical reality. Along the line, it is shown that significant progresses can be made by redefining, reinterpreting and reformulating those quantities.
Unlike the discussions in mathematics, the language, scope, perspective and objective of this paper are 
mostly physics-oriented.  Physicists and applied mathematicians should be able to find something
fundamental and interesting.

\section{\label{sec:level2}Trickiness of multi-variable limit}

In the conceptual sense the multi-variable limit is not much different from the single-variable limit, but defining and using multi-variable limits in reality is much trickier than people usually think.
In this section, we provide a concise overview of the subject. 

Let's start with an expression looking like an ordinary $0/0$ three-variable limit:
\beq \label{limit} \lim\limits_{(x,y,z)\to(0,0,0)} \frac{x^2+\sin(3y^2)+z^2}{x^2+y^2}.  \eeq
The behavior of Eq.~(\ref{limit}) around the limit point $(x,y,z)=(0,0,0)$ can be analyzed via the following if-then tests.  
If  $y$ and $z$ tend to zeros much faster than $x$, then the expression approaches   $1$. 
If $x$ and $y$ vary under the constraint $x=y$, while $z$ goes to zero much faster than $x$ and $y$, then the expression approaches $2$.  
However, if $z$ is nonzero and varies slowly while $x$ and $y$ tend to zeros rather fast, then the expression approaches $\infty$. These statements show that Eq.~(\ref{limit}) is multi-valued and cannot be treated as a definitely defined multi-variable limit.  (Similarly behaved single-variable ``limits'' exist, but their bad behavior is usually directly observable.)

In mathematics (calculus), to prevent using such equivocally defined ``limits'', there is a rule as follows\cite{courant, thomas}:   
a function's limit can be deemed as legitimately defined if and only if the function approaches  a unique (definite) value
no matter in what way the function's variables get close to the limit point.
Under the rule, the legitimacy of a single- or multi-variable ``limit'' is routinely tested by inspecting whether
the value of the ``limit'' has path-dependence (the path here refers to a line or a curve or a dotted curve in the variable space). Evidently, the path-dependence testing procedure and the aforementioned if-then testing procedure are essentially equivalent. 

However, the if-then testing procedure has its own merits and demerits. To see this is the case, let's first familiarize ourselves with a few simple concepts below.
 
Consider, for instance, a three-variable function $f(x,y,z)$ whose limiting behavior around a limit point $(a,b,c)$ is of interest. 
Then, the expression  
\beq \label{limit_0} f(\act  x\rangle_a,\act  y\rangle_b,\act  z\rangle_c) \equiv \lim\limits_{(x,y,z)\to (a,b,c)} f(x,y,z)\eeq
will be called a limit-like expression (rather than a limit). The word ``limit-like''  stresses the fact that there is a fairly high possibility that the expression is illegitimately defined (ill-defined or equivocally defined). 
 
Each of the following expressions will be referred to as  a conditional sublimit (or an ad hoc sublimit) of the limit-like expression  $f(\act  x\rangle_a,\act  y\rangle_b,\act  z\rangle_c)$:  
\ba
 &{\hspace{-0.2cm}}f(\act  x\rfloor,\act  y\rangle_b,\lfloor z\rangle_c),\,\;f(\lfloor x\rangle_a,\act  y\rfloor,\act z\rangle_c ),\,\; f(\act  x\rangle_a,\act  y\rangle_b,\act z\rfloor),\nonumber\\
 &{\hspace{-4.25cm}} f(\act  x{\&}y\rangle_{(a,b)},\lfloor z\rfloor ),\; \,{\rm and }\;  \cdots,
  \ea 
where 
the variables enclosed in $\act\cdot\cdot\rangle$ represent the ones whose  limiting processes are taken care of  immediately, the variables enclosed in $\lfloor \cdot\cdot\rfloor $ 
represent the ones whose limiting processes are frozen temporarily (or permanently), the variables connected by ``\&'' stand for the ones that behave collectively under certain constraints. In other words, each conditional sublimit is linked to a set of provisos that stipulate how the involved limiting processes are dealt with.

According to this terminology, if $f$ is defined by Eq.~(\ref{limit}), the $f$ is a limit-like expression having a large number of conditional sublimits. A further observation is that among  those conditional  
limits of the $f$ there are plenty that can be easily evaluated as if they are  single-variable limits. For example,
\ba
 \label{1.1} &&\hspace{-0.9cm}f(\lfloor x^2\rfloor ,\act  y^2\rangle_0,\act  z^2\rangle_0)\to {x^2}/{x^2} \to 1, \\
  \label{1.2} &&\hspace{-0.9cm}f(\act x^2\rangle_0, \lfloor y^2\rfloor ,\act  z^2\rangle_0)\to {\sin (3y^2)}/{y^2} \to 3, \\
\label{1.3} &&\hspace{-0.9cm}f(\act  x^2+ y^2\le \rho^2\rangle_{\rho\to 0},\lfloor z^2\rfloor )\to{z^2}/{0} \to \infty, \\
 \label{1.4}  &&\hspace{-0.9cm}f(\lfloor x=\kappa y\rfloor , \act z^2\rangle_0)\to \frac{(ky)^2+\sin(3y^2)}{(ky)^2+y^2}\to
 \frac{\kappa^2+3}{\kappa^2+1}. \ea  
This is to say, the limiting behavior of a limit-like quantity can be examined by evaluating its conditional sublimits, in particular those that are single-variable-like. 

It is now evident that a limit-like expression must be ill-defined if one of the following three criteria is met: i. The expression has two differently-valued conditional sublimits. ii. The expression has one $\infty$-valued conditional sublimit. (Herein, $\infty$ is regarded as an illegitimate limit since $\infty$ is not a definite value and no measurable quantity in physics is $\infty$-valued.) iii. The expression has one conditional sublimit whose value is constraint-dependent. 
Thus and so, we can, from Eqs.~(\ref{1.1}) and  (\ref{1.2}) or from  Eq.~(\ref{1.3}) alone or from Eq.~(\ref{1.4}) alone,  infer that expression~(\ref{limit}) is ill-defined. 

[Interested readers may use criterion iii to simply show the nonexistence of 
$\lim_{(x,y)\to (0,0)} {xy^2}/({x^2+y^4})$.]

A routine in the classroom is that if a limit-like expression is identified as ill-defined the ``limit'' is labeled as nonexistent and the activity related to it terminates utterly. 
However, there exists 
an informal and rather delicate practice:
after a limit-like quantity meets with some kind of difficulty, 
one (or more) of its conditional sublimits is defined and consequently applied as a conditional  substitute. Though the practice is pragmatic in many situations, it may, if used carelessly, become some kind of pitfall.
Illustrative examples will be given when
necessary and possible. All these tell us that, the introduction of the conditional sublimit is of more use than expected, and the application of the conditional sublimit needs more caution than expected.

There is another important point worth mentioning.
Quite often, the trouble is not that people lack skill to identify ill-defined quantities, but that there exist limit-like quantities involving limiting processes quite vaguely, and people  do not think of them as ``officially'' defined multi-variable limits let alone carefully inspect whether, or in what situations, they are ill-defined. 
To see how this can possibly happen, let's go through the following two commonplace examples.
  
As the first example, suppose that there is a two-variable quantity $g$ defined by 
\beq \label{gxy} g(x,y)= {1}/({x^2y^2+1}) \eeq
and the task given to us is to examine the behavior of $g$ around and along the $y$-axis. To the casual eye, 
this quantity behaves sometimes like an ordinary limit (when x gets close to the y-axis) and sometimes 
like an ordinary function (when y varies from one value to another); there seems no need to treat it
as a regular two-variable limit. However, a careful analysis shows otherwise.

It is easy to see that
\beq\label{heu1} g(\act  x\rangle_0,y) \equiv \lim\limits_{x\to 0} g(x,y)= 1\;({\rm with}\; y\;{\rm relatively}\; \rm inactive),\eeq
which is of use in certain contexts. But, can we thus claim that  $g(x,y)$ is  an invariant around and along the $y$-axis? 
At first glance, the claim is mathematically provable.  
Denoting $g(\act  x\rangle_0,y)$ as $h(y)$,  we obtain,  from Eq.~(\ref{heu1}), $h(y)=1$ and $h(y+\epsilon)=1$. According to the common knowledge in calculus, we arrive at
\beq \label{dhy} dh/dy=0,\eeq
which is a synonym for saying that $h(y)$ is an invariant. 
In this way, we have seemingly proven that $g(x,y)$  is an invariant around and along the $y$-axis.

\begin{figure}[ht]
\includegraphics[trim = 84.0mm 150.5mm 46mm 97mm, clip, width=8.65cm]{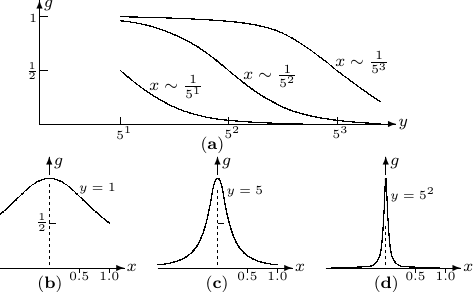}
\caption {{\bf (a)} $g=1/(x^2y^2+1)$ versus $y$ for different $x$; {\bf (b)} {\bf (c)}
and {\bf (d)} $g=1/(x^2y^2+1)$ versus $x$ for different $y$.}
\label{fig1}
\end{figure}

Is the above proof rigorous? The answer is no. In fact, there are two ways to refute it.
Firstly, let's look at how the proof treats its own proviso.
Note that although $h(y)$ in Eq.~(\ref{dhy}) seems like a single-variable function of $y$, it is not. The expression $h(y)$, or $g(\act  x\rangle_0, y)$ in Eq.~(\ref{heu1}), is actually a conditional sublimit, which should be expressed by the symbol $g(\act x\rangle_0,\lfloor y\rfloor )$ and needs to be linked with the specific proviso that the process of $\act x\rangle_0$ prevails over any possible variation of $y$.  When the proof concluded its conclusion, that specific proviso had been ignored conveniently but illogically. 
Secondly, let's examine whether $g$ is multi-valued around and along the $y$-axis. Note that the proviso that $\act x\rangle_0$ prevails over any variation of $y$ is not the only proviso that we can possibly  adopt. In fact, if we let $x$ be nonzero and vary slowly, and let $y$ become bigger and bigger rather rapidly, the value of $g$ approaches zero. This behavior is shown in Fig.~\ref{fig1}{\bf a}, and can be symbolically expressed by     
\beq\label{heu2} g(\lfloor x\rfloor ,\act  y\rangle_\infty)\to 0. \eeq
Eqs.~(\ref{heu1}) and (\ref{heu2}) tell us that the concerned behavior of $g$  should be described by       
 \beq\label{heu0} g(\act  x\rangle_0,\act  y\rangle_\infty)\to {\rm indefinable}. \eeq
Namely,  the behavior of $g$ around and along the $y$-axis involves two possible processes: $x\to 0$ and $y\to\infty$.
If and only if we consciously and consistently adopt a proviso under that either of the two processes prevails over the other, the concerned behavior of $g$ is well-defined. Otherwise, it is equivocally-defined (ill-defined).

The above discussion can be viewed from a different perspective. By noting that when $y$ becomes bigger and bigger the concerned $g(x,y)$  is like an  
ever-increasingly discontinuous function,
as shown in Figs.~\ref{fig1}{\bf b}, \ref{fig1}{\bf c}  and \ref{fig1}{\bf d},
we may realize that the limiting behavior of $g$ just unveiled is the indication that defining a function's limit is a risky business in a region where the function's value  involves a certain  type of discontinuity (embodying the relation between limits and continuity).

As the second example, let's recall how the concept of three-dimensional (3D) volume density evolves in our mind.
At an early stage of the education, everyone of us was taught that the volume density could be defined  by $d=m/U$ with $m$ and $U$ being the mass and volume of a concerned substance respectively. After we took calculus-level courses, we came to realize that the volume density should be redefined as  $d=\delta m/(\delta U)$ with $\delta m$ representing the mass of the substance enclosed in a sufficiently small volume element $\delta U$. Note that under the term ``sufficiently small'' there lurks a practically workable but subtle strategy of physicists. On one hand, physicists want $d=\delta m/(\delta U)$
to be qualified as  a $0/0$ limit so that $\delta m=d\cdot \delta U$ or $m=\int d\cdot dU$ can be applied rather freely (with no need to specify the size and shape of $\delta U$ or $dU$). On the other hand, physicists do not want $\delta U$, or $dU$, to be too small such that
the continuity assumption of the substance becomes invalid. (Using statistical arguments to uphold the needed continuity is more appropriate, but that is another story.) 

\begin{figure}[ht]
\includegraphics[trim = 57.0mm 266.5mm 94.9mm 20.5mm, clip, width=8cm]{f1}
\caption{Illustration of how the shape of $\delta U$ can affect the volume density $\delta m/(\delta U)$ for a 2D mass surface $S$.}
\label{fig0}
\end{figure}

Still, our concept about the volume density is vulnerable if we 
have only positive experience in using $d=\de m/\de U$. 
To be more specific, consider a 2D mass surface $S$ in 
Fig.~\ref{fig0}. Since the thickness of $S$ is negligible,  
$d=\delta m/(\delta U)$, if applied to $S$, takes an arbitrary value from $0$ to $\infty$, depending on the size and shape of $\delta U$ (each shape of $\delta U$ represents a particular constraint imposed on $\delta U$). In view of this arbitrariness, it should be acknowledged that the 3D volume density is ill-defined for a 2D mass (referred to as dimension mismatch hereafter). 

Though the concepts in this section are kind of trivial, they will play  
nontrivial roles in the rest of this paper.
\section{\label{sec:level3}Scattering cross sections in different reference frames}

To deal with particle collisions,  
a couple of $0/0$ limit-like quantities, called the scattering cross sections, 
are defined and consequently applied in non-equilibrium statistical 
mechanics\cite{liou,nolte,lerner,reif,dorf}. Oddly enough, one of them is ill-defined in terms of  its major application.

For simplicity, the discussion herein is on the understanding that all involved  particles are of the same mass, size and shape, but still distinguishable (possible in classical mechanics).  It is also assumed that 
the concerned particle interaction (collision) is short-ranged, one-to-one, and subject to the energy-momenta conservation law. 
Also, the molecular chaos hypothesis holds whenever a gas system is in the consideration.

 Although the usual textbook treatment starts with the particles of a beam scattered by a target particle at rest, the true concern therein is about beam-to-beam particle collisions. 
 In what follows, we 
 shall directly deal with beam-to-beam particle collisions.
 
Consider the setup depicted in Fig.~\ref{2beams}{\bf a}, where two narrow pure particle beams move towards each other.  (Herein, a particle beam is called a pure beam if all the particles in it possess exactly the same velocity. Another assumption with a pure beam is that the particles' positions in it are completely randomized.) In this paper, the following conventions are adopted almost always (unless mentioned otherwise):  
i. The particles belonging to the beam with velocity ${\bf v}$ are named as incident particles; and the particles belonging to the beam with velocity $\bf w$ as target particles. ii. The region where the concerned collisions take place is denoted as $d   U$ or $d^3{\bf r}$. 
 iii. After a collision between two particles with velocities $\bf v$ and $\bf w$ takes place,  the velocities of the two resultant particles are denoted by ${\bf v}'$ and ${\bf w}'$ respectively. 

\begin{figure}[ht]
\includegraphics[trim = 117.0mm 255.0mm 26.4mm 18.2mm, clip, width=8.65cm]{f1}\caption{{\bf (a)} The particles of two pure beams collide in a small region $d   U$; and  {\bf (b)} a pair of the collision-resultant particles are shown in the velocity space.}
\label{2beams}
\end{figure}

The velocity of the center-of-mass and the velocity of the incident particle  relative to the center-of-mass can then be defined as, respectively, 
\beq\label{cu} {\bf c}\equiv ({\bf v}+{\bf w})/2 \quad{\rm and} \quad {\bf u}\equiv {\bf v}-{\bf c}=({\bf v}-{\bf w})/2;\eeq
after $({\bf v},{\bf w}\to {\bf v}',{\bf w}')$, $\bf c$ and $\bf u$ become, respectively, 
\beq \label{cuprime}{\bf c}'\equiv({{\bf v}'}+{\bf w}')/2 \quad{\rm and} \quad {\bf u}'\equiv({{\bf v}'}-{\bf w}')/2.\eeq
Since our concern is only with one-to-one elastic collisions, the conservation law of energy and momenta reads:
\beq\label{em} {\bf c}'={\bf c} \quad{\rm and}\quad |{\bf u}'|=|{\bf u}|\equiv u.\eeq
It is worth noting that for a definite velocity pair $({\bf v},{\bf w})$ there exist infinitely many  velocity pairs $({{\bf v}'}, {\bf w}')$, and the paired ${{\bf v}'}$ and ${\bf w}'$ fall symmetrically onto two halves of the spherical shell labeled as the energy-momenta shell (EMS) in Fig.~\ref{2beams}{\bf b}. The reason why the shell is 2D lies in that there exist 4 energy-momenta equations serving as the constraints upon the 6  unknown components of ${{\bf v}'}$ and $\bf w'$. 
    
Based on these concepts and notations, the cross section called the scattering cross section in the center-of-mass reference frame takes the form 
\beq\label{sigma_cmf}  \sigma({\bf u},{\bf u}') = \sigma (\Omega)=\frac {d   N} {d   \Omega}=u^2\frac{d   N}{d  S} ,\eeq
in which the solid angle \( \Omega\) is defined by the direction of ${\bf u}'$ with respect to 
the original ${\bf u}$, $d\Omega$ is an infinitesimal solid-angle element  about $\Omega$,  
$d S$ is the infinitesimal surface element on the EMS subtending $d   \Omega$, and
 \(d   N\) represents the number of the incident particles that emerge, after the collisions, within $d   \Omega$, or on $d S$, per unit incident flux
($2un_{\bf v}=1$), unit target ($n_{{\bf w}}|d   U|=1$), and unit time (with $n_{\bf v}$ and $n_{\bf w}$ being the particle densities of the incident and target beams respectively). The reason why $\sigma$ is called the cross section in the center-of-mass frame is that   Eq.~(\ref{sigma_cmf}) is given explicitly in the frame. 

 Since  the collision-resultant particles distribute on the EMS  continuously (in the statistical sense), the value of $\sigma$ will be independent of how $dS$, or $d\Omega$, shrinks to the infinitesimal one. This means that $\sigma$ in
Eq.~(\ref{sigma_cmf}) is a well-defined two-variable $0/0$ limit. For this reason, we shall use  
 $dN=\sigma d\Omega=\sigma dS/u^2  $ rather freely. 
 
Next, consider another scattering cross section, which is herein called the cross section  in the laboratory frame (reflecting the fact that all quantities in its defining equation are given in the laboratory reference frame). Denoting it as $\lsigma$ (instead of $\sigma$), we have, according to the textbook treatment\cite{reif,kubo},
\beq\label{sigma_lf} \lsigma({\bf v},{\bf w}\to {{\bf v}'},{\bf w}')=\frac {d   N} {d  ^3{{\bf v}'} d  ^3{\bf w}'  }\eeq
where $d  ^3{{\bf v}'}\equiv d v_{x}' d   v_{y}' d   v_{z}'$ is a small velocity element about ${{\bf v}'}$, $d  ^3{\bf w}'\equiv d   w_{x}' d   w_{y}'d   w_{z}'$ is a small velocity element about ${\bf w}'$, and \(d   N\) is 
the number of the incident particles scattered into \(d  ^3{{\bf v}'}\) per unit incident flux,  unit target and unit time (while the involved target particles fall into $d  ^3{\bf w}'$ supposedly). In such treatment $\lsigma$ is defined and applied without specifying the size, shape and spatial orientation of $d^3{{\bf v}'}$, let alone those of $d  ^3{{\bf w}'}$, meaning that $\lsigma$ is indeed regarded as a legitimate $0/0$ limit. 

For the purpose of this paper, we now inspect whether or not $\lsigma$ 
can be regarded as a legitimate limit.

Even a simple contrast between Eq.~(\ref{sigma_cmf}) and Eq.~(\ref{sigma_lf}) provides us with crucial information.
Eq.~(\ref{sigma_cmf}) is based on the assumption that  all collision-resultant particles distribute on the two-dimensional EMS in the velocity space;  whereas, Eq.~(\ref{sigma_lf}) is based on the assumption  that each of ${\bf v}'$ and ${\bf w}'$ has a three-dimensional velocity space to fall into.
According to the last section, this should be termed as ``dimension mismatch''.
  
To have a more conclusive judgment, let's investigate whether  $\lsigma$ is multi-valued. As a limit-like expression,  Eq.~(\ref{sigma_lf}) has a conditional sublimit in the form 
\beq\label{sigma1} \lsigma(\act  d  ^3{{\bf v}'}\rangle_0, \lfloor d  ^3{\bf w}'\rfloor )=\lim\limits_{{d^3{{\bf v}'}}\to 0}\frac{dN}{d^3{{\bf v}'} d^3{{\bf w}'}}, \eeq
where $d^3{{\bf v}'}$ stands for a spherical ball in the velocity space with an ever-shrinking radius $\rho$ and $d^3{{\bf w}'}$ represents a spherical ball having a definite volume $dW'$. Provided that the centers of the two balls lie symmetrically on the EMS shown in Fig.~\ref{2beams}{\bf b}, then
 the value of Eq.~(\ref{sigma1}) is
\beq\label{sigma2} \lsigma(\act  d  ^3{{\bf v}'}\rangle_0, \lfloor d  ^3{\bf w}'\rfloor ) = \frac{   \sigma\pi \rho^2 /u^2}{(4\pi \rho^3/3) dW'  }=\frac {3\sigma}{4\rho u^2d W'},\eeq
in which $dN=\sigma dS/u^2=\sigma \pi\rho^2/u^2 $ has been used. 
However, if we let $d^3{{\bf v}'}$ stand for a small cube whose side $b$ is ever-shrinking and whose top and bottom are parallel with the EMS locally, we obtain, in otherwise the same context,
\beq\label{sigma20} \lsigma(\act  d  ^3{{\bf v}'}\rangle_0, \lfloor d  ^3{\bf w}'\rfloor ) = \frac{   \sigma b^2 /u^2}{b^3\cdot dW' } 
= \frac \sigma{bu^2dW'}. \eeq 
The above two expressions inform us that the value of $\lsigma$ highly depends on the size and shape of $d^3{{\bf v}'}$ (not to mention those of $d^3{{\bf w}'}$); and, as an additional observation, if $\rho\to 0$ or $b\to 0$, $\lsigma$ is $\infty$-valued. All these facts confirm that $\lsigma$ is ill-defined.

It should be remarked that the unveiled illegitimacy of $\lsigma$ has its root in the basic handling of physicists. As mentioned already, all the involved approaches treat particle-to-particle collisions as beam-to-beam particle collisions. This seemingly minor changeover gives us important advantages,
but the trade-off is that the particles emerging from beam-to-beam particle collisions distribute only on 2D surfaces in the velocity space. If we use an infinitesimal 3D velocity volume element to ``collect'' those particles,  the dimension mismatch difficulty will certainly arise.  (The difficulty can be eliminated if a finite 3D velocity volume element is used instead.)  
 
 In the textbook treatment,  the collision reversibility of particles is expressed by
\beq\label{vivi} \lsigma({\bf v},{\bf w}\to {{\bf v}'},{{\bf w}'})=\lsigma({{\bf v}'},{{\bf w}'}\to {\bf v},{\bf w}).\eeq
 But, according to the discussion above, neither side of Eq.~(\ref{vivi}) represents a definitely-valued physical quantity. [In other words, Eq.~(\ref{vivi})  is ``unfalsifiable''.]

People may still want a valid expression for collision reversibility.
 An investigation of Eq.~(\ref{sigma_cmf}) tells us that the cross section $\sigma$ can be rewritten as
\beq \label{new_sigma} \sigma(\Omega_{\bf uu'})=u^2\frac{d   N}{d S_{{\bf v}'}}\equiv
\sigma({\bf v},{\bf w}\to d   S_{{\bf v}'},d S_{{\bf w}'}), \eeq
in which $d S_{{\bf v}'}$ and $d S_{{\bf w}'}$ stand for two identical infinitesimal 2D patches lying symmetrically on the EMS (allowing $\bf v'$ and $\bf w'$ to fall onto respectively). 
With help of this new expression of $\sigma$, we obtain, formally in the laboratory reference frame,  
\beq \label{sigsig}\sigma({\bf v},{\bf w}\to d   S_{{\bf v}'},d   S_{{\bf w}'})=\sigma({{\bf v}'},{{\bf w}'}\to d   S_{\bf v},d   S_{\bf w}).\eeq
The validity of Eq.~(\ref{sigsig}) can be justified by noticing the fact that $\sigma(\Omega_{\bf uu'})= \sigma(\Omega_{\bf u'u})$.

Eq.~(\ref{sigsig}), though error-free in the mathematical sense, provides no help in terms of constructing the standard Boltzmann scattering operator. 


\section{\label{sec:level4}The scattering-out rate from a phase volume element}

Another $0/0$ limit-like quantity, called the scattering-out rate, is investigated in this section. It turns out that the scattering-out rate can be, and should be, viewed from a different standpoint.
 
In the existing kinetic theory\cite{liou,nolte,lerner,reif}, 
 the distribution function (sometimes called the probability density) of an ideal dilute gas is governed by the Boltzmann equation
\beq\label{be_eu} R_t=R^{\bf v}+R^{\bf F}+R^{\rm s.in}-R^{\rm s.out},\eeq
in which $R_t\equiv (\pt f/\pt t)_{{\bf r},{\bf v}}$ is the change rate of the distribution function at $({\bf r},{\bf v})\equiv (x,y,z,v_x,v_y, v_z)$, $R^{\bf v}=-{\bf v}\cdot (\pt f/\pt {\bf r})$ describes  how the particles are driven by the velocity $\bf v$, $R^{\bf F}=-{\bf F}/m\cdot (\pt f/\pt {\bf v})$ describes the influence of the macroscopic force ${\bf F}\equiv (F_x,F_y,F_z)$, and $R^{\rm s.in}-R^{\rm s.out}$  
stands for the Boltzmann scattering operator with $R^{\rm s.in}$ and $R^{\rm s.out}$ being   called the scattering-in rate and the scattering-out rate respectively.   

Although different approaches derive the Boltzmann equation somewhat differently, 
the operator 
 $R^{\rm s.in}-R^{\rm s.out}$ is unanimously formulated  by  
examining how particle collisions drive particles into, and out of, a fixed small phase volume element $d\beta\equiv d^3{\bf r}d^3{\bf v}$. This type of approach is usually called the Eulerian approach since the chosen control volume is a Eulerian-type control volume (fixed relative to the coordinate system). In this section, we shall strictly follow the Eulerian approach until meeting with insurmountable difficulties. 

In the Eulerian approach, the change rate of the local distribution function $R_t$  is actually defined by  
\beq\label{rt_def} R_t= \lim\limits_{d t\to 0,d\beta\to 0} \frac {(d N)_{dt, d\beta}}{dt d\beta}=\left.\frac{\pt f}{\pt t}\right|_{{\bf r},{\bf v}} ,\eeq
where 
$(d N)_{dt,d\beta}$ is the net increment of the particle number in $d\beta$ during $dt$ (from the start time to the end time). Eq.~(\ref{rt_def}) informs us that  $R_t$ is a limit-like quantity and the legitimacy of it needs to be carefully inspected, as  stressed in Sect.~\ref{sec:level2}. In fact,  every other term in Eq.~(\ref{be_eu}) needs to be inspected in a similar manner.

To make our inspection less burdensome, consider the following assumptions: i. ${\bf F}=0$. ii. The  $x$-axis is set along the direction of the concerned $\bf v$ so that  $v_y=v_z=0$. And, iii. the incoming particle beams are so arranged that there will be no collision-resultant particles whose final velocity is exactly equal to $\bf v$, and thus  $R^{\rm s.in}=0$ (dealing with $R^{\rm s.in}$ is purposely avoided in this section). Under these simplifications, Eq.~(\ref{be_eu}) is reduced to 
\beq \label{dddd}R_t=R^{v_x}-R^{\rm s.out},\eeq
in which $R^{v_x}= {[{(dN)^{v_x\rm .in}-(dN)^{v_x\rm .out}}]}/{(dtd\beta)}$  will be called the fluid term and $R^{\rm s.out}={(dN)^{\rm s.out} }/({dtd\beta})$  
will be called the kinetic term. Concerning these two terms, there are two tasks in front of us. The first is to examine whether each of the two terms is well-defined, and the second is to examine whether the events related to the two terms are mutually exclusive (demanded by the addition rule of probability).

The fluid term $R^{v_x}$ can be formulated in terms of fluid mechanics. With reference to Fig.~\ref{fig3}{\bf a}, we have:
\begin{align}\label{boundary} 
 (d   N)^{v_x\rm .in}-(d   N)^{v_x\rm .out}&= v_xd   t d   yd   zd  ^3{\bf v} ( f^l-f^r)\notag \\
&=- v_x \frac {\pt f}{\pt x} d   t d  \beta ,\end{align} 
where $f^l$ and $f^r$ are the distribution functions at the left and right ends of $d x$ respectively. We thus obtain 
\beq\label{rvx} 
R^{v_x}=\frac {(d   N)^{v_x\rm .in}-(d   N)^{v_x\rm .out}}{d   t d  \beta}=-v_x\frac {\pt f}{\pt x}.\eeq
This result is independent of how $dt$ and $d\beta$ shrink to their zeros,
and hence $R^{v_x}$ is a legitimate $0/0$ limit.

\begin{figure}[ht]
\includegraphics[trim = 74.2mm 102mm 44.0mm 164.2mm, clip, width=8.7cm]
{f1}\caption{{\bf (a)} How the particles of $f_{\bf v}$ move in, and out of, $d\beta$ due to $v_xdt$. {\bf (b)} How the particles of $f_{\bf v}$ are scattered out of $d\beta$ by $f_{\bf w}$. {\bf (c)} The combination of these two effects.}
\label{fig3}
\end{figure}

The kinetic term $R^{\rm s.out}$ can be formulated with help of Fig.~\ref{fig3}{\bf b}, in which the  concerned collision region is filled with a number of letter x (symbolizing the collisions therein). As said before, we here refer to the particles belonging initially to $f_{\bf v}$ and $f_{\bf w}$ as the incident particles and the target particles respectively. 
Thus, the number of the incident particles scattered into $d\Omega$ due to the collisions  in $d^3{\bf r}$ during $dt$ is, by virtue of Eq.~(\ref{sigma_cmf}),
\beq\label{dn} (d N)^{d \Omega} = \sigma (d \Omega ) \cdot (2uf_{\bf v}d^3{\bf v})\cdot(f_{\bf w}d^3{\bf w}d^3{\bf r} )\cdot(d t),\eeq
in which the incident flux $2u f_{\bf v}d^3{\bf v}$, the target particle number $f_{\bf w}d^3{\bf w}d^3{\bf r}$, and the time interval $dt$ emerge explicitly since each of them is no longer equal to unit.  
The total number of the scattering-out particles is
\beq\label{out00} (d N)^{\rm s.out}= d^3{\bf r}d^3{\bf v} d t \int 2u \sigma  f_{\bf v}  f_{\bf w}d^3{\bf w}  d\Omega ,\eeq
in which  $\int f_{\bf w}d^3{\bf w}$ stands for all the particle beams coming to $d^3{\bf r}$ 
to collide with the beam $f_{\bf v}d^3{\bf v}$.
Finally, we obtain 
\beq\label{out0}R^{\rm s.out}= \lim\frac {(d N)^{\rm s.out}}{dt d^3{\bf r} d^3{\bf v}}
= \int  2u\sigma f_{\bf v} f_{{\bf w}} d^3{\bf w}  d\Omega. \eeq
For those who know the existing kinetic theory, the above derivation of $R^{\rm s.out}$ has nothing new. In view of that Eq.~(\ref{out0}) needs  no information about how $d^3{\bf r}$,  
$d^3{\bf v}$ and $dt$ shrink to their zeros,  we see that the resultant $R^{\rm s.out}$ is in itself a well-defined $0/0$ limit.

So far, everything is fully consistent with that in the textbook treatment. 

Once we start inspecting whether or not the events related to $R^{v_x}$ and $R^{\rm s.out}$ are mutually exclusive, unexpected things pop up.
 In Fig.~\ref{fig3}{\bf c}, the events related to $R^{v_x}$ and $R^{\rm s.out}$ are depicted in a combined way. Evidently, there is a double counting when $(dN)^{v_x\rm .out}$
and $(dN)^{\rm s.out}$ are taken into account simultaneously. This simply means that expression (\ref{dddd}) violates the addition rule for probability. 

In fact, we can devise a virtual experiment to directly observe the consequence of  the double counting.
Suppose that in the setup  shown in Fig.~\ref{fig3}{\bf c} the particle beam represented by $f_{\bf v}$ passes through $d^3{\bf r}$ constantly (before and after $dt$) and the colliding beam represented by $f_{\bf w}$ is applied to $d^3{\bf r}$  just after the start time of $dt$.  By evaluating the particle number in $d^3{\bf r}d^3{\bf v}$ at the start time of $dt$ and 
the particle number in $d^3{\bf r}d^3{\bf v}$ at the end time of $dt$, it is found that  
$R_t\equiv \lim d N/(dtd\beta)$  defined by Eq.~(\ref{rt_def}) is equal to
\beq\label{rt01} R_t (\lfloor v_xdt =\kappa dx\rfloor, \act dydzd^3{\bf v}\rangle_0)= (-1+\frac \kappa 2)R^{\rm s.out},\eeq
where 
 $\kappa\equiv v_xdt/(dx)\leq 1$. 
In physics, Eq.~(\ref{rt01}) implies that the value of $R_t$ depends on the experimental parameter $\kappa$. In mathematics, Eq.~(\ref{rt01}) implies that 
$R_t$ is an ill-defined ``limit'' (see criterion iii in Sect.~\ref{sec:level2}). 

It is now tenable to argue that the problem just revealed is unavoidable for the Eulerian approach. As has been stated, the
  goal of the Eulerian approach is to formulate the net increment of the particle number inside a fixed infinitesimal $\de\beta$ during an infinitesimal $dt$. To achieve the goal, the working hypothesis is that all the concerned particles can be distinguished into a number of distinctive groups, of which each is linked to 
 a particular driving-out or driving-in mechanism. Under the working hypothesis, $R_t\equiv \lim d N/(dtd\beta)$ becomes      
\beq\label{brief}
\frac{-(dN)^{v_x\rm .out}+(dN)^{v_x\rm .in}-\cdots-(dN)^{\rm s.out}+(dN)^{\rm s.in}}{dt d\beta}. \eeq
If we let $dt$ be truly short, each term in the numerator of expression (\ref{brief}) is rather small and can be evaluated independently. 
However, if we let $d\beta$ be too small, Eq.~(\ref{brief}) ceases to make sense. To see why,   think of the following scenario: if $dx$ of $d\beta$ is much smaller than $v_xdt$, all the particles in $d\beta$, including those initially in it and those entering it during $dt$, will move out of $d\beta$ instantly (in a time much shorter than $dt$), making the regular meaning of $(dN)^{v_x\rm .in}$,$\cdots$, $(dN)^{\rm s.out}$ or $(dN)^{\rm s.in}$ groundless.
This will be called the distinguishing difficulty hereafter.

One question arises: Can we interpret $R^{\rm s.out}$ 
without the involvement of the smallness of $d\beta$ and $dt$? 
Interestingly, the question has a positive answer. 
If we look at the case from the perspective of the Lagrangian-type path approach (in which a material volume element along a path is considered as a control volume), we find  that
the collision frequency therein is independent of the smallness of $d\beta$ and $dt$.
More interestingly, by adopting the new perspective, we gain not one but two advantages. The first is that the only driving-out mechanism we need to consider is the one that has been formulated by Eq.~(\ref{out0}). The second is that  Eq.~(\ref{out0}) can be integrated easily and meaningfully. To put the second advantage in context, we plug Eq.~(\ref{out0}) into the Boltzmann equation (\ref{be_eu}), and obtain (with $R^{\rm s.in}$ disregarded again)
\beq\label{dn11} \left.\frac {d   f_{\bf v}} {d   t}\right|_l=v\frac {d   f_{\bf v}} {d   l}=-f_{\bf v} \int d^3{\bf w} d\Omega 2u\sigma  f_{{\bf w}},\eeq
where $l$ represents the path length of the concerned particles and $v\equiv |{\bf v}|$ is the local speed of these particles.  
This formula is essentially identical to another formula in the literature\cite{reif}:
\beq \label{dn22}\frac {dp}{pdt}=v\frac {dp}{pdl}=-\tau^{-1}= -\int d^3{\bf w} d\Omega 2u\sigma  f_{\bf w} ,\eeq 
in which $p$ is the survival probability of a test particle and $\tau$ is the 
average collision time (or relaxation time). 

The integration of  Eq.~(\ref{dn22}) over a finite path $\delta l$ gives us the path-survival probability of a moving particle
\beq\label{path_p}p(\delta   l) =\exp \left( -\int_{\delta   l} \frac{dl}{|\bf v|} \int d^3{\bf w}d\Omega 2u\sigma f_{\bf w}\right).\eeq 
This formula is quite meaningful in terms of describing how a particle source gives contribution to the distribution function elsewhere; and it should be included as an indispensable part of any would-be kinetic theory\cite{chen1}. 

The discussion in this section has shown that the scattering-out rate formulated for the Boltzmann equation makes a better  sense in the path-approach. Namely, the scattering-out rate from a fixed phase volume element is not a good concept, but the scattering-out rate from a path is. 

\section{\label{sec:level5}The distribution function of collision-resultant particles}

In this section, we explore in what context the beam-to-beam  particle collisions can be  formulated. It turns out that the concerned distribution function, as another $0/0$ limit, has to be redefined.

\begin{figure}[ht]
\includegraphics[trim = 67.5mm 214mm 63.5mm 55.1mm, clip, width=8.7cm]{f1}\caption{({\bf  a}) Two gases $f_{{\bf v}'}$ and $f_{{\bf w}'}$ colliding in the region $\Delta U$. ({\bf  b})  Concerning the resultant distribution function, two different analyses give two different predictions. }
\label{pra11}
\end{figure}

Let's first review what the existing approach has to say. 
For future convenience, consider the situation depicted in Fig.~\ref{pra11}{\bf a}, where   
 two time-independent colliding gases represented by  
 $f_{{\bf v}'}$ 
 and 
 $f_{{\bf w}'}$ 
 are naturally separated from the collision-resultant gases represented by $f_{\bf v}$ 
and $f_{\bf w}$ (only $f_{\bf v}$ is shown in the figure).
It should be noted that in this section we consider the collision $({{\bf v}'},{{\bf w}'})\to ({\bf v},{\bf w})$, instead of the collision $({\bf v},{\bf w})\to ({{\bf v}'},{{\bf w}'})$. 

A basic, but often ignored, concept in the existing approach is the scattering-in rate 
defined by:
\beq\label{s_in} R^{\rm s.in}=\lim\limits_{dt\to 0, d^3{\bf r}\to 0,d^3{\bf v}\to 0}\frac {(d N)^{\rm s.in}} {d
 t d^3{\bf r}d^3{{\bf v}}}, \eeq
in which $(dN)^{\rm s.in}$ is the number of the particles that initially belong to $f_{{\bf v}'}$  and are later scattered into $d^3{\bf v}$ by the collisions occurring in $d^3{\bf r}$ during $dt$.
With help of the collision reversibility expressed by Eq.~(\ref{vivi}), this rate becomes (details omitted for brevity)
 \beq\label{s_in1}  R^{\rm s.in}=\int 2u\sigma f_{{\bf v}'}f_{{\bf w}'}d^3{\bf w}d\Omega. \eeq
Eventually, Eq.~(\ref{s_in1}) enters into the Boltzmann equation as a part of the change rate of the local distribution function.

The above formulation, though existing for a long time, suffers from the conceptual difficulties listed below: 
 \begin{itemize}
 \item
As analyzed in Sect.~\ref{sec:level3} and some of our previous papers\cite{chen1,chen2}, no 3D infinitesimal velocity volume element is allowed to collect the particles emerging from beam-to-beam particle collisions, otherwise the ``dimension mismatch'' difficulty will certainly arise. In other words, Eqs.~(\ref{s_in}) and (\ref{s_in1}), as well as Eq.~(\ref{vivi}), are invalid in the first place.
\item
 If we compute the number of the particles that enter a fixed infinitesimal $d^3{\bf r}d^3{\bf v}$ during an infinitesimal $dt$, $d^3{\bf r}$ cannot be too small; otherwise all the particles, including those initially in $d^3{\bf r}$ and those entering $d^3{\bf r}$ during $dt$, will move out of $d^3{\bf r}$ instantly (before the end of $dt$). The two conflicting requirements
 that $d^3{\bf r}d^3{\bf v}$ should be infinitesimal and  $d^3{\bf r}$ should not be too small have been construed as the ``distinguishing difficulty'' in Sect.~\ref{sec:level4}. 
\item
 In the existing theory,  $R_t-R^{\bf v}-R^{\bf F}=0$ is sometimes called the collisionless Boltzmann equation, and 
adding $R^{\rm s.in}-R^{\rm s.out}$ to the right side of it gives us the collisional Boltzmann equation. If we examine the collisionless Boltzmann equation we find that there is a good symmetry in terms of 
$\pt/\pt{\bf r}$ and $\pt/\pt{\bf v}$. But, if we examine $R^{\rm s.in}$ and $R^{\rm s.out}$  we find that all operations in $R^{\rm s.in}$ or $R^{\rm s.out}$  are performed 
in the ${\bf v}$-space (with $\bf r$ serving merely as a parameter). There is no good explanation for this unsymmetry\cite{chen3}. 
Another symmetry-related puzzle is that while the the collisionless Boltzmann equation requires, and yields, full differentiability of distribution function (except on the boundary), the operator $R^{\rm s.in}-R^{\rm s.out}$
does not require, nor yield, continuity of distribution function. 
\end{itemize}

Despite of all these conceptual difficulties, it is still fair to say that the existing theory has captured some substantial features of particle collisions, for why else does the theory yield so many ``great'' results in the literature. To find out what should be retained and what should be discarded, we now use the theory to compute the situation given in Fig.~\ref{pra11}{\bf a}. Hopefully, the computation will uncover something practically important. 
  
Suppose that the positive $x$-axis is set along the symmetry axis OA of the collision region $\Delta U$ (with O as the origin of the $x$-axis)
and that the concerned velocity $\bf v$ is also along the OA-axis (the $x$-axis). Then, the Boltzmann equation (\ref{be_eu}) on the $x$-axis is 
\beq\label{fx0}
  v_x\frac 
  {\pt f_{\bf v}(x)}{\pt x}=\int 2u\sigma f_{{\bf v}'}f_{{\bf w}'}d^3{\bf w}d\Omega,\eeq
in which $v_y=v_z=0$, ${\bf F}=0$ and $(\pt f_{\bf v}/\pt t)=0$  have been used; and higher-order collisional effects (proportional to $\sigma^2, \sigma^3,\cdots$) have been neglected. 
We eventually obtain  
\beq\label{fx}  f_{\bf v}(x)  =\frac{1}{v_x}\int^{x}_0 {dx'}\int 2u\sigma f_{{\bf v}'}f_{{\bf w}'} 
d^3{\bf w}d\Omega.\eeq
This expression can be numerically treated, and the result is shown by the dotted curve in Fig.~\ref{pra11}{\bf b}.

Yet, the reality seems to tell us a different story.  Suppose that  we are an observer moving  backwards away from the collision region (say along the x-axis). What shall we observe at different places?
Given the fact that the collision region acts increasingly like a point-size particle source, 
it is reasonable to believe that the observed distribution function is governed by the inverse-square law
at large distances, and the solid curve in Fig.~\ref{pra11}{\bf b} is the one describing the true behavior of $f_{\bf v}(x)$. 

Though the above analysis denies the accuracy of Eq.~(\ref{fx}), we can still learn a lot by interpreting Eq.~(\ref{fx}) from different perspectives. Suggestions we may possibly get include: i. The particle collisions at a place can serve as a particle source to give contribution to the distribution function elsewhere. ii. For an observer at a position point, the concerned distribution function can be formulated by integrating all the contributions from the upstream sources (with no need to consider the downstream sources). iii. The concerned distribution function shouldn't be defined as 
$ (dN)/{d^3{\bf r}d^3{\bf v}}$ in which both $d^3{\bf r}$ and $d^3{\bf v}$ are infinitely small. This is due to the fact that if both $d^3{\bf r}$ and $ d^3{\bf v}$ are infinitely small, we 
have no choice but to integrate the contribution from a ``one-dimensional'' upstream path that  has zero volume and contains no colliding particles.

Surprisingly, if we carefully take these suggestions into consideration, almost all the 
aforementioned conceptual difficulties evaporate.

 \begin{figure}[ht]
\includegraphics[trim = 65mm 119.5mm 73.0mm 152.5mm, clip, width=8.7cm]{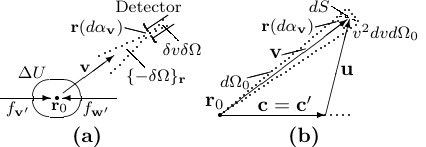}
\caption{Illustration of the collision-resultant particles detected by a detector, which is shown {\bf (a)} in the position space and {\bf (b)} in the velocity space.}
\label{collision_illu}
\end{figure}

Let's now consider Fig.~\ref{collision_illu}{\bf a}, which is just a portion  of Fig.~\ref{pra11}{\bf a}.
For conceptual visibility, suppose that a virtual particle detector with an infinitesimal inlet is placed at $\bf r$ right towards the collision region ($\bf r$ is located outside or inside the collision region but not necessarily on the OA axis). Furthermore, suppose that the detector can  detect only the particles whose velocity magnitude is within a small but finite 
 range $\delta v$ and whose velocity direction is within a small but finite solid-angle element $\delta \Omega$.    
Thus,  the following ``distribution function'' is the distribution function ``observed'' by the 
detector:
\beq\label{dis_fun} \bar f(\act  d^3{\bf  r}\rangle_0,\delta   v,\delta \Omega)\equiv\frac{1}{
{\bar v}^2 \de   v \de \Omega}\lim\limits_{dtd\alpha_{\bf v}\to 0}\int_{\delta  v\delta \Omega}
\frac{dN}{vdt  d \alpha_{\bf v} },\eeq 
where \(\act  d  ^3{\bf  r}\rangle_0\) represents an infinitesimal position volume element at the  inlet, 
\(\bar v^2\equiv (v_h^2+v_lv_h+v_l^2)/3\) with \(v_h\) and \(v_l\) being the highest and lowest speeds of \(\de v\), \(d \alpha_{\bf v}\equiv d\alpha\cdot |\cos\theta|\) with $d\alpha$ being the 
infinitesimal area of the inlet and $\theta$ being the angle between the normal of $d\alpha$ and the velocity of the concerned entering particles, and \(dN\) is the particle number recorded by the detector during \(dt\).

The quantity $\bar f$ in Eq.~(\ref{dis_fun}) can be taken as a conditional sublimit associated with the proviso that $d^3{\bf r}$ is infinitesimal and $\bar v^2 \de v\de \Omega$ is a small but finite velocity volume element (whose size and shape are fixed). In other words, although $\bar f$ is 
  an ordinary distribution function in the position space, it is an average distribution function 
 over the velocity volume element $(v_h^3-v_l^3)\cdot \delta \Omega/3$.
 
By examining  the upstream paths of the detector in Fig.~\ref{collision_illu}{\bf a}, it is found that 
only the collisions taking place in
the spatial cone \(\{-\delta  \Omega\}_{\bf  r}\) can possibly give direct contribution to \(\bar f\). 
For that reason, we call 
the region enclosed by \(\{-\delta  \Omega\}_{\bf  r}\) the effective  zone.
By denoting  a specific point in the effective zone as \({\bf  r}_0\), and denoting the solid-angle element formed by $d\alpha_{\bf v}$ and ${\bf r}-{\bf r}_0$ as $d\Omega_0$, we have, in reference to Fig.~\ref{collision_illu}{\bf b}, 
\beq\label{omega_0} d\Omega_0= d \alpha_{\bf v} /|{\bf  r}-{\bf  r}_0|^2.\eeq
Since $d\Omega_0$ is infinitesimal and  \(\delta  \Omega\) is finite, 
any particle that starts from the effective zone and enters the detector at a speed within \(\delta v\)  is  qualified as a particle belonging to \(dN\) in Eq.~(\ref{dis_fun}).

Hence, the contribution  to \(dN\) due to the collisions in \(d^3{\bf  r}_0\)  can be written as, with help of Eq.~(\ref{dn}),
\beq\label{dn_0}  \int_{\delta  v}\int_{v^2dvd\Omega_0}\sigma\cdot(dS/u^2) 2uf_{{\bf  v}'}d^3{\bf  v}' f_{{\bf  w}'} d^3{\bf  w}'d^3{\bf  r}_0 dt  ,\eeq
where \(dS\) is the surface element lying on the EMS defined by ${\bf  c}={\bf c}'$ and $|{\bf u}|=|{\bf u}'|$, and enclosed by
\(v^2dvd\Omega_0\), as shown in Fig~\ref{collision_illu}{\bf b}.

To truly integrate Eq.~(\ref{dn_0}), we make use of the variable transformation
\(d^3{\bf  v}'d^3{\bf  w}'\to d^3{\bf  c}'d^3{\bf  u}'\), where \({\bf  c}'=({{\bf v}'}+{{\bf w}'})/2\) and \({\bf  u}'=({{\bf v}'}-{{\bf w}'})/2\), and obtain
\beq \int_{\delta v}\int_{v^2dvd\Omega_0} \sigma\cdot(dS/u^2) 2uJf_{{\bf  v}'} f_{{\bf  w}'}  d^3{\bf  c}'d^3{\bf  u}' d^3{\bf  r}_0dt  ,\eeq
in which \(J\) is the Jacobian 
\beq J=\left\| \frac {\pt({\bf  v}',{\bf  w}')}{\pt({\bf  c}',{\bf  u}')}\right\| 
=8.\eeq
Noticing that
\beq \int d^3{\bf  u}'=\int u'^2du'd\Omega'=\int d\Omega'\int u^2du,\eeq
where \(\Omega'\) is the direction of \({\bf  u}'\), and that
\beq \int_{v^2dvd\Omega_0} du dS=v^2dv d\Omega_0=\frac{v^2dv d \alpha_{\bf v} }{|{\bf  r}-{\bf  r}_0|^2},\eeq
we obtain, from Eq.~(\ref{dis_fun}),
\beq\label{result_f} \bar f=\frac{1}{
{\bar v}^2 \delta  v\delta   \Omega}\int d^3{\bf  r}_0 dv d^3{\bf  c}'d\Omega' \frac{16uv\sigma}{|{\bf  r}-{\bf  r}_0|^2}f_{{\bf v}'}f_{{\bf w}'},\eeq
where \(\int d^3{\bf  r}_0\) is over the effective zone \(\{-\delta   \Omega\}_{\bf  r}\), \(\int dv\) is over \(\delta   v\), \(\int d^3{\bf  c}'\) is over the entire velocity space, and \(\int d\Omega'\) is over \(0-4\pi\).
It should also be noted that all the other quantities in this formula, such as \(\sigma\), \(f_{{\bf  v}'}\) and \(f_{{\bf  w}'}\), are defined with help of \({\bf  u}\) and \({\bf  u}'\), in which \({\bf  u}={\bf  v}-{\bf  c}={\bf  v}-{\bf  c}'\) with \({\bf  v}=v({\bf  r}-{\bf  r}_0)/|{\bf  r}-{\bf  r}_0|\), and \({\bf  u}'\) is defined by \(|{\bf  u}|\) and the direction of \(\Omega'\).

Finally, a few remarks about Eq.~(\ref{result_f}): i. If $|{\bf r}-{\bf r}_0|$ is 
much larger than the size of the collision region, the value of the result  obeys the inverse square law (just as expected). ii. With slight modifications, the formulation is applicable to many other situations\cite{chen3}. For instance, if $f_{{\bf v}'}$ and $f_{{\bf w}'}$ 
are time-dependent, the only modification needed is a time-shift. Namely, we are supposed to use 
 $f_{{\bf v}'}(t-|{\bf r}-{\bf r}_0|/v)$ and $f_{{\bf w}'}(t-|{\bf r}-{\bf r}_0|/v)$ to replace $f_{{\bf v}'}$ and $f_{{\bf w}'}$ respectively.
   iii. The integral in the result involves an infinitely large region in the position space and an infinitely large region in the velocity space, meaning that the result is in principle discontinuous with respect to $\bf r$ and to $\bf v$. 
   

\section{\label{sec:level6}Different interpretations of the distribution function along a path}

Eq.~(\ref{dn11})  in Sect.~\ref{sec:level4} can be interpreted as saying that the distribution function of  a collsionless gas is
invariant along a particle's path; whereas Eq.~(\ref{result_f}) in Sect.~\ref{sec:level5} can be interpreted as saying that  the distribution function caused by a finite-size particle source obeys the inverse-square law at large distances.
Are these two statements in conflict with each other? If they are, can we reconcile them
in some sense? 
To answer these interesting and fundamental questions, we here investigate the basic framework of the existing kinetic theory.
 
 One of the standard ways to derive the existing kinetic theory is through the following  four-level Liouville-theorem formalism\cite{reif,schwabl}:
\newline\noindent
1. Liouville's theorem (path-constancy of phase volume). \newline\noindent
2. Path-invariance of distribution function (deemed as a corollary of Liouville's theorem in many publications).\newline\noindent
3.  The collisionless Boltzmann equation.\newline\noindent
4.  The collisional Boltzmann equation.\newline\noindent
Although the collisional Boltzmann equation received some criticisms in the history of physics, today's physics community accepts the above 4-level formalism with almost no resistance. 

Another way to derive the Boltzmann equation is via the so-called fluid-mechanics treatment\cite{reif,dorf}, in which the involved phase space is divided into many small phase volume elements, and then how the particles get out of, or get into, these
 volume elements  is examined and formulated. In terms of this treatment, the boundary and initial conditions of the Boltzmann equation are assumed to be given in advance. 

The mainstream consensus is that these two derivations are essentially equivalent. Another common conception is that  
if a kinetic computation needs to be conducted (in today's or tomorrow's computer), the concepts and methodology should be consistent with those in the fluid-mechanics treatment. 

As can be noticed, the preceding sections of this paper have already challenged the  theory described above.
 For one thing, 
 Sect.~\ref{sec:level2} suggested that if a ``path-invariant'' is established with help of limiting processes the legitimacy of the invariant needs to be inspected carefully. 
For another, 
Sects.~\ref{sec:level3}, \ref{sec:level4} and \ref{sec:level5} demonstrated that, in terms of including the collisional effects, the existing theory is far from perfect, to say the least. 

As a matter of fact, even a cursory contrast between the path-approach (referring to level 1 and 2 of the Liouville-theorem formalism) and the equation-approach (referring to level 3 and 4 of the formalism)
 gives us   
something to worry about. 
Suppose the job we need to do is to formulate the time behavior of a gas that suffers only scarce particle-particle collisions.
At first sight, either the path-approach alone or the equation-approach alone could be used to do the job,  and the end results would  be basically the same.  
However, if we delve deeply into the subject, we might realize that these two approaches are too different to yield basically the same results.
The most noticeable difference between the two is that, throughout the path-approach the path-information of individual particle plays an obvious and indispensable role,
whereas in the equation-approach the Boltzmann equation and the given initial and boundary conditions 
form a complete set, leaving almost no role for the path-information to play. 
And, as a related point, in the path-approach what happens on the boundaries is determined mostly by the initial condition and the path information, whereas in the equation-approach the boundary condition is supposed to be given in advance. (Moreover, it is easy to find that the two approaches are quite different in terms of including and dealing with discontinuity.)

With all these questions in mind, we now inspect every ambiguity in the Liouville-theorem
formalism. For simplicity, the inspection will be based on what happens with collisionless and force-free ideal gases.

\begin{figure}[ht]
\includegraphics[trim = 63.7mm 225.0mm 67mm 43.3mm, clip, width=8.7cm]{f2}
\caption{Illustration of how a material volume element moves and evolves in the $x$-$v_x$ space.}
\label{free_motion2}
\end{figure}

The path-constancy of phase volume (level 1) can be illustrated by a figure very much  like
 Fig.~\ref{free_motion2} (only in the $x$-$v_x$ plane for technical reasons), in which a material phase volume element, denoted as $\db\equiv \delta x\delta v_x$, is moving and evolving with its volume (area) unchanged. If we wish to use Fig.~\ref{free_motion2} to justify the path-invariance of distribution function (level 2), the following three assumptions need to be satisfied: i. At the initial time the particles in $\db$ distribute continuously (in the statistical sense),
 so the initial distribution function is well-defined.  ii. When $\db$ evolves in phase space, the 
distribution function in $\db$ keeps well-defined (as will be clarified gradually, this is only conditionally true). iii. The particles in $\db$ are not created or destroyed.
 The textbook treatment\cite{landau,walter,huang,schwabl,harris} takes all these assumptions for granted, and asserts that  the distribution function of a collisionless gas behaves like an incompressible fluid and the
path-invariance of distribution function is a corollary of Liouville's theorem. 
 
\begin{figure}[ht]
\includegraphics[trim = 82.2mm 188.5mm 52.2mm 84.5mm, clip, width=8.7cm]{f2}
\caption{An alternative view on how a distribution function evolve in the $x$-$v_x$ phase space.}
\label{free_motion}
\end{figure}  

Interestingly, there exist schematic figures capable of challenging the above reasoning.  Fig.~\ref{free_motion} is one of them, in which the initial distribution function $f_0(x,v_x)$ has been sliced into many thin slices according to their different $v_x$. Thanks to the fact that each of the slices keeps moving with its own speed and along its own path, these slices, as a whole, will constantly spread (expand) in the position dimension and  constantly contract in the velocity dimension. 
If a tiny creature is attached to a slice's peak, his/her eyes, as a physical instrument, will see  fewer and fewer particles around him/her. Eventually, only 0 percent of the initial particles can be seen by the creature.

\begin{figure}[ht]\includegraphics[trim =  66mm 615mm 476.4mm 310.5mm, clip, width=8.6cm]{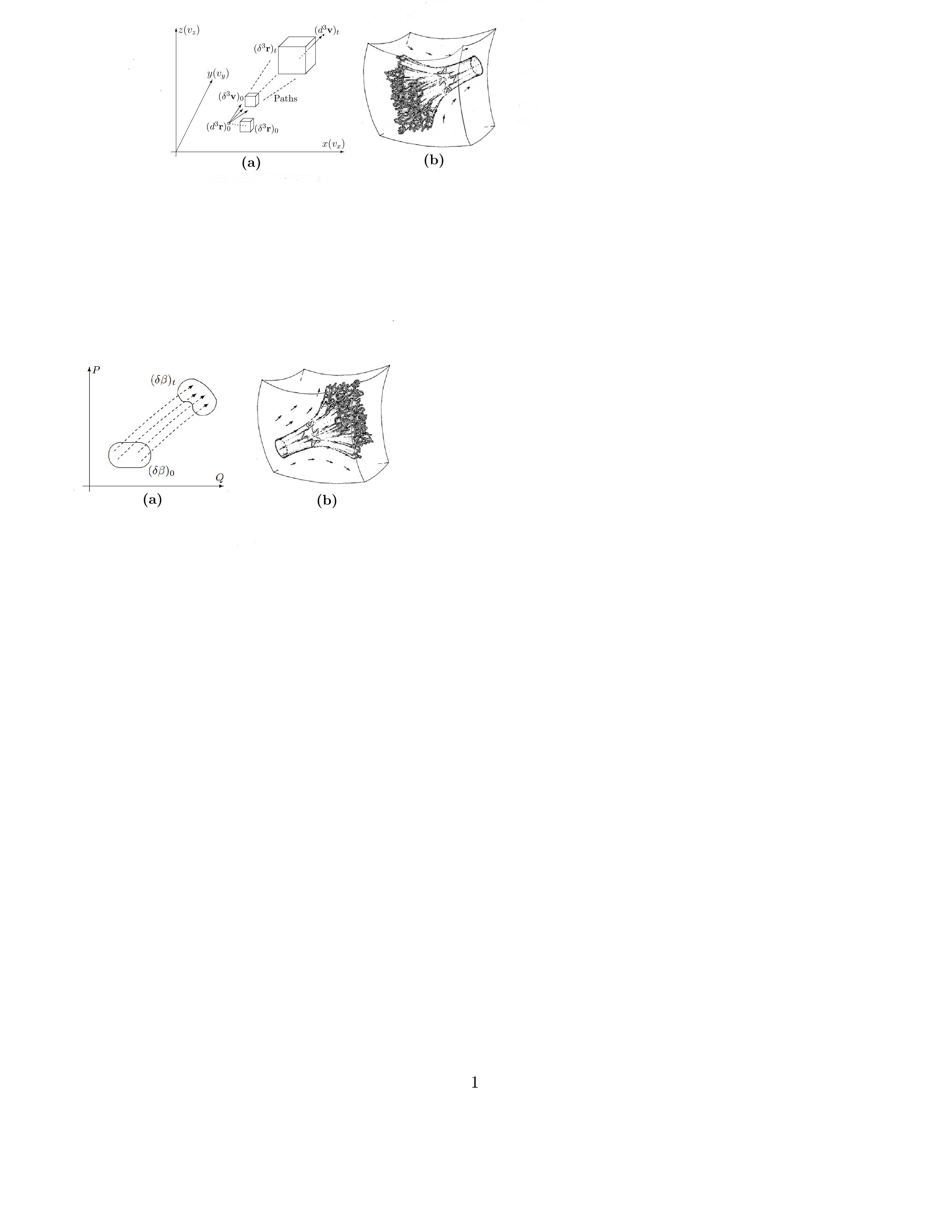}
\caption{The evolution of a phase volume element depicted {\bf (a)} in some textbooks, and {\bf (b)} on {\it physicstravelguide.com}.}
\label{liou1}
\end{figure}

At this stage, it is of significance to visit some related graphs in the existing publications.  
In textbook-style books\cite{harris,walter}, Fig.~\ref{liou1}{\bf a}, or something like that, is given to 
illustrate Liouville's theorem (or its corollary). If this figure governs one's mind, one will  
willingly agree with the analogy between  a phase volume element in phase space and  an incompressible fluid in the daily life.  
However, on {\it physicstravelguide.com}, a metaphorical figure, Fig.~\ref{liou1}{\bf b}, is given to exhibit the evolution of a phase volume element. If that figure dominates one's mind, one may likely assume that a phase volume element evolves like an ever expanding ``sponge'' (how to measure its density is a problem worth discussing). 
All these mean that the high-dimensional phase space is inherently counter-intuitive, and 
visualizing the behavior of  a gas in it is more sophisticated than we think.

\begin{figure}[ht]
\includegraphics[trim = 134.0mm 900.5mm 377.4mm 21.0mm, clip, width=8.6cm]{liou_11.png}
\caption{{\bf (a)} A position element and a velocity element evolve independently and collaboratively. {\bf (b)} 
Metaphorical illustration of the genesis of a phase volume element.  }
\label{liou2}\end{figure}

To comprehend the behavior of a 6D phase volume element more directly, let's look at Fig.~\ref{liou2}{\bf a}, in which a small but finite position element $(\delta^3{\bf r})_0$ and a small but finite velocity element $(\delta^3{\bf v})_0$ evolve independently and collaboratively. We first concern ourselves with $(d^3{\bf r})_0$, which is just a tiny  part of $(\delta^3{\bf r})_0$. It is easy to see that $(d^3{\bf r})_0$ will, along the different paths defined by different velocities in
$(\de^3{\bf v})_0$, spread into an ever-expanding position element 
 $(\delta^3{\bf r})_t$. According to Liouville's theorem, we have $|(d^3{\bf r})_0(\delta^3{\bf v})_0|=|(\delta^3{\bf r})_t(d^3{\bf v})_t|$, where $(d^3{\bf v})_t$ represents a tiny ever-contracting velocity element in the velocity space defined at each and every single position point in $(\delta^3{\bf r})_t$. In reference to Fig.~\ref{liou2}{\bf a}, the theorem further shows us that the expansion of $(\de^3{\bf r})_t$ is constant and uniform [$|(\de^3{\bf r})_t|\propto t^3$], and the contraction of 
$(d^3{\bf v})_t$ is constant and rather drastic [$|(d^3{\bf v})_t| \propto t^{-3}$].
 As for $(\delta^3{\bf r})_0(\delta^3{\bf v})_0$, their ever expanding-contracting behavior is about the same (except that the expansion and contraction are less intense). 
With the realization that  we can associate  
the expansion in Fig.~\ref{liou1}{\bf b} with what happens in the position subspace 
and associate the 
contraction in Fig.~\ref{liou1}{\bf b} with what happens in the velocity subspace, we should be able to conclude that Fig.~\ref{liou1}{\bf b}, rather than Fig.~\ref{liou1}{\bf a}, is the adequate illustration of an ever-evolving phase volume element.

Incidentally, if we wish to know how a phase volume element $(\db)_0\equiv (\delta^3{\bf r})_0(\delta^3{\bf v})_0$ is formed in phase space, we need to do the time-reversal from $t_0=0$ to the time $t<0$ in a figure similar to Fig.~\ref{liou2}{\bf a}. By doing just that, we shall find that, just as Fig.~\ref{liou1}{\bf b} can adequately describe the development of a phase volume element, Fig.~\ref{liou2}{\bf b} can adequately describe the genesis of a phase volume element. (In Sect.~\ref{sec:level5}, the concept of effective zone was introduced by the same token.) 

Then, the key issue of this section becomes as to how to define the distribution function along and around a particle's path  for the particles confined in such ever expanding-contracting phase volume elements.

Conventionally, the distribution function along a particle's path is loosely defined  by:
\beq\label{hatf} \hat f(\delta \beta ,l)=\frac{\delta N(\delta \beta,l)}{|\delta \beta|}=\frac{\delta N(\delta^3{\bf r}\delta^3{\bf v},l)}{|\delta^3{\bf r}\delta^3{\bf v}|},\eeq
where  $l$ represents the path length of  the particle ($l=0$ at the initial time)
and $\delta\beta\equiv\delta^3{\bf r}\delta^3{\bf v} $ stands for a ``rather small'' phase volume element whose ``center'' moves along $l$, 
$|\delta\beta|$ is the volume of 
$\delta\beta$ and $\delta N$ is the number of the particles possibly found in $\db$. 
By applying Eq.~(\ref{hatf}) to different situations, such as those given in
Figs.~\ref{liou1}{\bf b} and \ref{liou2}{\bf b}, it is found that  Eq.~(\ref{hatf}) is as equivocally-defined as Eq.~(\ref{gxy}) in Sect.~\ref{sec:level2}. The similarities between the two are the following: 
i. Like Eq.~(\ref{gxy}), Eq.~(\ref{hatf}) behaves sometimes like an ordinary limit, and 
sometimes like an ordinary function.
ii. Like Eq.~(\ref{gxy}), Eq.~(\ref{hatf}) ``defines'' an  invariant with help of a limiting process.
iii. Like Eq.~(\ref{gxy}), Eq.~(\ref{hatf}) needs to deal with an ever-increasingly  discontinuous
function along a path (which can be seen in Figs. \ref{free_motion} and \ref{liou1}{\bf b}). 
iv. Like Eq.~(\ref{gxy}), Eq.~(\ref{hatf}) involves two possible processes
$\db\to 0$ and $l\to \infty$. If there is a proviso under that one of the two processes surely prevails over the other, it is 
well-defined, otherwise it is ill-defined.  
A somewhat distinctive thing related to Eq.~(\ref{hatf}) is that people from different physics communities tend to impose different provisos upon it  (sometimes unconsciously).

For physicists who view kinetic theory mostly from the theoretical perspective, the phase element $\db\equiv \de^3{\bf r}\de^3{\bf v}$, if used to define a distribution function, will almost always approach zero in an overwhelming manner. Under this mindset, ``the distribution function along a path'' is nothing but a conditional sublimit of Eq.~(\ref{hatf}): 
\beq\label{f_def}f({\bf r},{\bf v})\equiv \lim\limits_{\db\to 0}
\frac {\de N(\db,l)}{|\db|}\; ({\rm with\;{\it l}\;\rm relatively \; inactive}),\eeq
where $({\bf r},{\bf v})$ represents the end phase point of the path $l$. 
For conceptual accuracy, we shall call $f$ defined by Eq.~(\ref{f_def}) the point-wise distribution function (since $\act \db\rangle_0$ eventually looks like an isolated point in phase space). 
Under this definition, in reference to Fig.~\ref{free_motion2}, we arrive at 
\beq\label{f100}  f({\bf r},{\bf v})=f_{l=0}({\bf r}_0,{\bf v}_0)\equiv f_{t=0}({\bf r}_0,{\bf v}_0),\eeq
in which $({\bf r}_0,{\bf v}_0)$ is the start phase point  of the path $l$.
Eq.~(\ref{f100}) is formally the same as  the path-invariance of distribution function in the existing theory. But, in order to ensure the usability of Eq.~(\ref{f100}), it is necessary to assume that $f_0({\bf r}_0,{\bf v}_0)$ is well-defined and the involved limiting process of $\act\db\rangle_0$ overwhelms the increase of $l$.

For physicists who are working on experimental and numerical studies or for those who wish to inspect  the full legitimacy of Eq.~(\ref{hatf}), imposing different provisos upon $\db\equiv \de^3{\bf r}\de^3{\bf v}$ is practically and theoretically imperative.   
To see what happens
when different provisos are introduced, consider a new function $\bar f$:
\beq\label{finitef} \bar f_{ \lfloor\de^3 {\bf v}\rfloor} (l)\equiv
\hat f(\lfloor \de^3{\bf r}\rangle_0, \lfloor\de^3 {\bf v}\rfloor, l) \eeq
in which $\hat f$ is defined by Eq.~(\ref{hatf}), and $\lfloor\de^3 {\bf v}\rfloor$ stands for a small velocity volume element with fixed size and shape. Or, consider another new function $\bar f$:
\beq\label{finitef10} \bar f_{ \lfloor\db\rfloor} (l)\equiv
\hat f( \lfloor\db \rfloor, l)\eeq  
in which $\lfloor\db\rfloor$ stands for a small phase volume element with fixed size and shape.  The distribution functions defined by these two expressions may be called the region-averaged distribution functions. 
For the situations shown in Fig.~\ref{free_motion} (or Fig.~\ref{free_motion2}, or Fig.~\ref{liou1}{\bf b}), we find that      
\beq\label{le_zero}\bar f_{ \lfloor\de^3 {\bf v}\rfloor}(l\to \infty)\to 0 \eeq
and 
\beq\label{le_zero2} \bar f_{ \lfloor\db \rfloor}(l\to \infty)\to 0. \eeq
In other words, if  examined from physical or numerical perspective, the distribution function does not have to be path-invariant (answering the questions raised at the beginning of this section).

In practice, we almost always deal with finite $\de^3{\bf r}\de^3{\bf v}$ 
and finite $l$, and hence neither Eq.~(\ref{f100}) nor Eq.~(\ref{le_zero2}) can be applied without reservation. At the end of this section, we shall deal with such collisionless gases in a more general approach.

In the theoretical sense, Eqs.~(\ref{f100}), (\ref{le_zero}) and (\ref{le_zero2}) 
can be collectively expressed by  
\beq\label{ff100} \hat f(\act \delta^3{\bf r}\rangle_0,\act \delta^3{\bf v}\rangle_0, \act l
\rangle_\infty)\to {\rm indefinable},\eeq
which implies that every gas system is destined to become a discontinuous one. 
In view of that this is a highly unconventional conclusion, we shall do more exploration about it. It turns out that the reality is even worse than what Eq.~(\ref{ff100}) can possibly express. 
 
Consider a collisionless and force-free gas given in  Fig.~\ref{shape}. Suppose that 
 the initial distribution function $f_0({\bf r}_0,{\bf v}_0)$ is well-defined and  differentiable at every point inside the region $\Sigma$, and  $f_0=0$ outside $\Sigma$.   
It is not difficult to see that if $({\bf r}_0,{\bf v}_0)\rightarrow ({\bf r},{\bf v})$ defines a particle's path from time $t_0$ to time $t$,
then $f_t({\bf r},{\bf v})$ is well-defined and differentiable unless $({\bf r}_0,{\bf v}_0)$ is a boundary point of $\Sigma$.
With help of $({\bf r}_0,{\bf v}_0)=({\bf r}-{\bf v}t,{\bf v})$ and Eq.~(\ref{f100}), we obtain    
\beq\label{f_ptx}  \left. \frac {\pt f_t}{\pt \br}\right|_{\br,\bv}=\left. \frac {\pt f_0} {\pt \br_0}\right|_{\br_0,\bv_{0}}{\hspace {2.45cm}}\eeq
and
\beq\label{f_ptv} \left. \frac {\pt f_t}{\pt \bv}\right|_{\br,\bv}=-t\left. \frac {\pt f_0} {\pt \br_0}\right|_{\br_0,\bv_{0}}+\left. \frac {\pt f_0} {\pt \bv_0}\right|_{\br_0,\bv_{0}}. \eeq 
Eqs.~(\ref{f100}), (\ref{f_ptx}), (\ref{f_ptv}) and Fig.~\ref{shape} are quite 
informative.
For one thing, 
Eq.~(\ref{f100}),  Eq.~(\ref{f_ptx}) and Fig.~\ref{shape}{\bf a} tell us that when the initial distribution function (as a source) has a definite profile in the position space, the distribution function at a later time
(as the image of the source) will have infinitely many same-shaped profiles
in the position space, one by one without distance in between, thanks to the fact that each $\bf v$ can carry a source profile to a new place. This is to say, the initial boundary will migrate to virtually everywhere in the position space.
 For another thing, Eq.~(\ref{f100}), Eq.~(\ref{f_ptv}) and Fig.~\ref{shape}{\bf b} inform us that when the source distribution function moves to another position point $\bf r$, a new almost same-shaped profile in the velocity space will arrive at $\bf r$, and  the sharpness of the new profile will be enhanced roughly by the factor $t=l/|{\bf v}|$ [Note that while the first term on the right side of Eq.~(\ref{f_ptv}) keeps growing and growing along the path, the second term keeps unchanged]. 

 \begin{figure}[ht]
\includegraphics[trim = 50.2mm 154.0mm 82.0mm 114.0mm, clip, width=8.6cm]{f2}
\caption{An initial distribution function produces a later distribution function consisting of:  {\bf (a)} many images in the position space, and  {\bf (b)} many images in the velocity spaces, of which each is defined at a position point.}
\label{shape}
\end{figure}  

By adopting the notion that a function's discontinuity can be characterized by its local $\infty$-differentiation, we can view the above discussion from the continuity perspective.  Eqs.~(\ref{f100}), (\ref{f_ptx}), (\ref{f_ptv}) and Fig.~\ref{shape} show  that if $f_0$ is discontinuous somewhere (due to the initial or boundary conditions), the discontinuity will migrate to virtually everywhere in the position space (which may be called the first type of ever-spreading discontinuity).
More than that, if $|\partial f_0/\partial \br_0|$ is fairly large somewhere, all the distribution functions spreading away from the place will become ever-increasingly discontinuous in the velocity spaces (which may be called the second type of ever-spreading discontinuity). 
Finally, a related observation  is that  although particle collisions can erase discontinuity along every path,  particle collisions can also create  discontinuity along every path, as indicated at the end of Sect.~\ref{sec:level5} (which may be regarded as the third type of ever-spreading discontinuity).

The most critical conclusion of the above discussion is that, in view of the ubiquitous ever-spreading and ever-evolving discontinuity in realistic gases, all the concepts, methodologies and theories solely based on continuity and differentiability need certain reconsideration.  

\begin{figure}[ht]
\includegraphics[trim = 105.5mm 73.2mm 31.5mm 194.8mm, clip, width=8.6cm]{f2}
\caption{Illustration of how the region-averaged distribution function at a time
can be determined by the initial distribution function.}
\label{example}
\end{figure}  

Before finishing this topic, it is worth mentioning that although the point-wise distribution function is not a generally valid concept, 
there are situations in which  the point-wise distribution function can provide help to determine the region-averaged distribution function.
As a specific  example, consider Fig.~\ref{example}, in which the initial distribution function $f_{t_0=0}$  of a 1D collisionless gas is well-defined  in the regions labeled 
 $(\delta x)_1$ and $(\de x)_2$ [provided $f_0=0$ outside $(\delta x)_1$ and $(\delta x)_2$]. Suppose our task is to determine the region-averaged distribution function $\bar f_t$ at a later time $t>t_0$ in the four finite phase volume elements ${(\db)_t^i}$, ${(\db)_t^{ii}}$,
${(\db)_t^{iii}}$ and ${(\db)_t^{iv}}$. 

Denote a phase point in the region ${(\db)_t^i}$ as $({\bf r},{\bf v})$.
Let $({\bf r}_0,{\bf v}_0)$ be the phase point defined by the path-reversal: 
\beq\label{time-r} ({\bf r},{\bf v})\xrightarrow{t\to t_0} ({\bf r}_0,{\bf v}_0). \eeq 
It is obvious that if $({\bf r}_0,{\bf v}_0)$ is not a boundary point of the initial distribution function,
then both $f_0({\bf r}_0,{\bf v}_0)$ and $f_t({\bf r},{\bf v})$ are well-defined.  Hence, 
\begin{align} \label{finitef1}
\bar f_t [{(\db)_t^i}]
&=\int_{(\db)_t^i} f_t({\bf r},{\bf v}) d^3{\bf r}d^3{\bf v} \Big/ |(\db)^i_t| \notag\\
&=\int_{(\db)_{0}^i} f_0({\bf r}_0,{\bf v}_0) d^3{\bf r}_{0}d^3{\bf v}_{0} \Big/ |(\db)^i_{0}|.
\end{align}  

In the same manner, $\bar f_t$  in ${(\db)_t^{ii}}$,
${(\db)_t^{iii}}$ and ${(\db)_t^{iv}}$ can be determined. 

It should be noted that the changeover of the integration region from $(\db)_t$ to $(\db)_0$ is in general quite troublesome. For instance, if 
$(\db)_t$ is a 6D cuboid, $(\db)_0$ will be  a ``sponge-like'' volume element in the phase space, as shown in Fig.~\ref{liou2}{\bf b}. For this reason, we should, in almost all practical cases,  use
 \beq
 \bar f_t[(\db)_t] =\int_{(\db)_t} f_0({\bf r}_0,{\bf v}_0) d^3{\bf r}d^3{\bf v}\Big/ |(\db)_t|,\eeq
in which $f_0({\bf r}_0,{\bf v}_0) $
is the initial distribution function (provided it is piece-wise well-defined) with $({\bf r}_0,{\bf v}_0)$ being defined by the path-reversal Eq.~(\ref{time-r}).
\section{Summary}
In this paper, we have shown that

\begin{itemize}
\item A number of limit-like quantities used in the existing non-equilibrium statistical mechanics are  actually ill-defined (equivocally-defined).

\item By redefining, reinterpreting and reformulating those ill-defined limits, significant progresses  can be made.

\item Generally speaking, continuously distributed particles and discontinuously distributed particles  coexist 
in a real gas. In this sense, both the point-wise distribution function and the region-averaged distribution function find their use in formulating the gas dynamics.

\item In terms of solving the distribution function from a given initial distribution function, getting help from some type of path-reversal approach is inevitable.

\end{itemize}
The concepts, methodologies and conclusions in this paper are fundamentally different from those in the conventional theory. Possible impact on fluid mechanics, turbulence studies, entropy studies,  plasma physics, quantum statistical mechanics and general  statistical approaches remains to be seen.

\section{Acknowledgments}
 The author is very grateful to Drs. O. Penrose, Robert G. Littlejohn, Lu Qishao,
  Guan Keying, Guo Hanying, 
Zhang Tianrong,  Ying Xingren, V. Travkin, W. Hoover and many others for their direct or indirect encouragement. Discussions with them have been pleasant and stimulating.

\bibliography{pcondi}

\providecommand{\noopsort}[1]{}\providecommand{\singleletter}[1]{#1}%
\begin{thebibliography}{20}%
\makeatletter
\providecommand \@ifxundefined [1]{%
 \@ifx{#1\undefined}
}%
\providecommand \@ifnum [1]{%
 \ifnum #1\expandafter \@firstoftwo
 \else \expandafter \@secondoftwo
 \fi
}%
\providecommand \@ifx [1]{%
 \ifx #1\expandafter \@firstoftwo
 \else \expandafter \@secondoftwo
 \fi
}%
\providecommand \natexlab [1]{#1}%
\providecommand \enquote  [1]{``#1''}%
\providecommand \bibnamefont  [1]{#1}%
\providecommand \bibfnamefont [1]{#1}%
\providecommand \citenamefont [1]{#1}%
\providecommand \href@noop [0]{\@secondoftwo}%
\providecommand \href [0]{\begingroup \@sanitize@url \@href}%
\providecommand \@href[1]{\@@startlink{#1}\@@href}%
\providecommand \@@href[1]{\endgroup#1\@@endlink}%
\providecommand \@sanitize@url [0]{\catcode `\\12\catcode `\$12\catcode
  `\&12\catcode `\#12\catcode `\^12\catcode `\_12\catcode `\%12\relax}%
\providecommand \@@startlink[1]{}%
\providecommand \@@endlink[0]{}%
\providecommand \url  [0]{\begingroup\@sanitize@url \@url }%
\providecommand \@url [1]{\endgroup\@href {#1}{\urlprefix }}%
\providecommand \urlprefix  [0]{URL }%
\providecommand \Eprint [0]{\href }%
\providecommand \doibase [0]{https://doi.org/}%
\providecommand \selectlanguage [0]{\@gobble}%
\providecommand \bibinfo  [0]{\@secondoftwo}%
\providecommand \bibfield  [0]{\@secondoftwo}%
\providecommand \translation [1]{[#1]}%
\providecommand \BibitemOpen [0]{}%
\providecommand \bibitemStop [0]{}%
\providecommand \bibitemNoStop [0]{.\EOS\space}%
\providecommand \EOS [0]{\spacefactor3000\relax}%
\providecommand \BibitemShut  [1]{\csname bibitem#1\endcsname}%
\let\auto@bib@innerbib\@empty
\bibitem [{\citenamefont {Gelbaum}\ and\ \citenamefont {Olmsted}(2003)}]{B1}%
  \BibitemOpen
  \bibfield  {author} {\bibinfo {author} {\bibfnamefont {B.~R.}\ \bibnamefont
  {Gelbaum}}\ and\ \bibinfo {author} {\bibfnamefont {J.~M.~H.}\ \bibnamefont
  {Olmsted}},\ }\href@noop {} {\emph {\bibinfo {title} {Counterexamples in
  Analysis}}}\ (\bibinfo  {publisher} {Courier Corporation},\ \bibinfo {year}
  {2003})\BibitemShut {NoStop}%
\bibitem [{\citenamefont {Steen}\ and\ \citenamefont {Jr}(1978)}]{B2}%
  \BibitemOpen
  \bibfield  {author} {\bibinfo {author} {\bibfnamefont {L.~A.}\ \bibnamefont
  {Steen}}\ and\ \bibinfo {author} {\bibfnamefont {J.~A.~S.}\ \bibnamefont
  {Jr}},\ }\href@noop {} {\emph {\bibinfo {title} {Counterexamples in
  Topology}}}\ (\bibinfo  {publisher} {Springer-Verlag New York Inc},\ \bibinfo
  {year} {1978})\BibitemShut {NoStop}%
\bibitem [{\citenamefont {Stoyanov}(2013)}]{B3}%
  \BibitemOpen
  \bibfield  {author} {\bibinfo {author} {\bibfnamefont {J.~M.}\ \bibnamefont
  {Stoyanov}},\ }\href@noop {} {\emph {\bibinfo {title} {Counterexamples in
  Probability, Third Edition}}}\ (\bibinfo  {publisher} {Dover Books on
  Mathematics},\ \bibinfo {year} {2013})\BibitemShut {NoStop}%
\bibitem [{\citenamefont {Wigner}(1959)}]{wigner}%
  \BibitemOpen
  \bibfield  {author} {\bibinfo {author} {\bibfnamefont {E.~P.}\ \bibnamefont
  {Wigner}},\ }\href@noop {} {\emph {\bibinfo {title} {The unreasonable
  effectiveness of mathematics in the natural sciences}}}\ (\bibinfo
  {publisher} {Richard Courant lecture in mathematical sciences},\ \bibinfo
  {year} {1959})\BibitemShut {NoStop}%
\bibitem [{\citenamefont {Courant}\ and\ \citenamefont {John}(1989)}]{courant}%
  \BibitemOpen
  \bibfield  {author} {\bibinfo {author} {\bibfnamefont {R.}~\bibnamefont
  {Courant}}\ and\ \bibinfo {author} {\bibfnamefont {F.}~\bibnamefont {John}},\
  }\href@noop {} {\emph {\bibinfo {title} {Introduction to Calculus and
  Analysis, Volume II/1}}}\ (\bibinfo  {publisher} {Springer-Verlag New York,
  inc.},\ \bibinfo {year} {1989})\BibitemShut {NoStop}%
\bibitem [{\citenamefont {Whittaker}\ and\ \citenamefont
  {Watson}(1996)}]{thomas}%
  \BibitemOpen
  \bibfield  {author} {\bibinfo {author} {\bibfnamefont {E.~T.}\ \bibnamefont
  {Whittaker}}\ and\ \bibinfo {author} {\bibfnamefont {G.~N.}\ \bibnamefont
  {Watson}},\ }\href@noop {} {\emph {\bibinfo {title} {Thomas' Calculus}}}\
  (\bibinfo  {publisher} {Cambridge University Press},\ \bibinfo {year}
  {1996})\BibitemShut {NoStop}%
\bibitem [{\citenamefont {Liouville}(1838)}]{liou}%
  \BibitemOpen
  \bibfield  {author} {\bibinfo {author} {\bibfnamefont {L.}~\bibnamefont
  {Liouville}},\ }\href@noop {} {\bibfield  {journal} {\bibinfo  {journal} {J.
  Math. Pures Appl.}\ }\textbf {\bibinfo {volume} {3}},\ \bibinfo {pages} {342}
  (\bibinfo {year} {1838})}\BibitemShut {NoStop}%
\bibitem [{\citenamefont {Nolte}(2010)}]{nolte}%
  \BibitemOpen
  \bibfield  {author} {\bibinfo {author} {\bibfnamefont {D.~D.}\ \bibnamefont
  {Nolte}},\ }\bibfield  {title} {\bibinfo {title} {The tangled tale of phase
  space},\ }\href@noop {} {\bibfield  {journal} {\bibinfo  {journal} {Physics
  today}\ }\textbf {\bibinfo {volume} {63(4)}},\ \bibinfo {pages} {33}
  (\bibinfo {year} {2010})}\BibitemShut {NoStop}%
\bibitem [{\citenamefont {Lerner}\ and\ \citenamefont {Trigg}(1991)}]{lerner}%
  \BibitemOpen
  \bibfield  {author} {\bibinfo {author} {\bibfnamefont {R.}~\bibnamefont
  {Lerner}}\ and\ \bibinfo {author} {\bibfnamefont {G.}~\bibnamefont {Trigg}},\
  }\href@noop {} {\emph {\bibinfo {title} {Encyclopaedia of Physics, 2nd
  Edition}}}\ (\bibinfo  {publisher} {VHC publishers},\ \bibinfo {year}
  {1991})\BibitemShut {NoStop}%
\bibitem [{\citenamefont {Reif}(1965)}]{reif}%
  \BibitemOpen
  \bibfield  {author} {\bibinfo {author} {\bibfnamefont {F.}~\bibnamefont
  {Reif}},\ }\href@noop {} {\emph {\bibinfo {title} {Fundamentals of
  Statistical and Thermal Physics}}}\ (\bibinfo  {publisher} {McGraw-Hill Book
  Company},\ \bibinfo {year} {1965})\BibitemShut {NoStop}%
\bibitem [{\citenamefont {Dorfman}(1999)}]{dorf}%
  \BibitemOpen
  \bibfield  {author} {\bibinfo {author} {\bibfnamefont {J.}~\bibnamefont
  {Dorfman}},\ }\href@noop {} {\emph {\bibinfo {title} {An Introduction to
  Chaos in Nonequilibrium Statistical mechanics}}}\ (\bibinfo  {publisher}
  {Cambridge University Press},\ \bibinfo {year} {1999})\BibitemShut {NoStop}%
\bibitem [{\citenamefont {R.~Kubo}\ and\ \citenamefont
  {Hashitsume}(1991)}]{kubo}%
  \BibitemOpen
  \bibfield  {author} {\bibinfo {author} {\bibfnamefont {M.~T.}\ \bibnamefont
  {R.~Kubo}}\ and\ \bibinfo {author} {\bibfnamefont {N.}~\bibnamefont
  {Hashitsume}},\ }\href@noop {} {\emph {\bibinfo {title} {Statistical
  Mechanics II, 2nd Edition}}}\ (\bibinfo  {publisher} {Springer-Verlag},\
  \bibinfo {year} {1991})\BibitemShut {NoStop}%
\bibitem [{\citenamefont {Chen}(2006)}]{chen1}%
  \BibitemOpen
  \bibfield  {author} {\bibinfo {author} {\bibfnamefont {C.}~\bibnamefont
  {Chen}},\ }\href@noop {} {} (\bibinfo {year} {1999-2006}),\ \Eprint
  {https://arxiv.org/abs/physics.gen-ph/0812.4343; cond-mat/0412396, /0504497,
  /0608712; physics/9908062, /0006009, /0006033, /0010015, /0305006, /0311120,
  /0312043} {physics.gen-ph/0812.4343; cond-mat/0412396, /0504497, /0608712;
  physics/9908062, /0006009, /0006033, /0010015, /0305006, /0311120, /0312043}
  \BibitemShut {NoStop}%
\bibitem [{\citenamefont {Chen}(2002)}]{chen2}%
  \BibitemOpen
  \bibfield  {author} {\bibinfo {author} {\bibfnamefont {C.}~\bibnamefont
  {Chen}},\ }\href@noop {} {\bibinfo {title} {Mathematical investigation of the
  boltzmann collisional operator}} (\bibinfo {year} {2002}),\ \bibinfo {note}
  {it was officially published on Il Nuovo Cimento B, V117B, p177},\ \Eprint
  {https://arxiv.org/abs/physics/0006038} {physics/0006038} \BibitemShut
  {NoStop}%
\bibitem [{\citenamefont {Chen}(2015)}]{chen3}%
  \BibitemOpen
  \bibfield  {author} {\bibinfo {author} {\bibfnamefont {C.}~\bibnamefont
  {Chen}},\ }\bibfield  {title} {\bibinfo {title} {An alternative approach to
  particle-particle collisions},\ }\href
  {https://doi.org/10.4236/jmp.2015.66082} {\bibfield  {journal} {\bibinfo
  {journal} {Journal of Modern Physics}\ }\textbf {\bibinfo {volume} {6}},\
  \bibinfo {pages} {772} (\bibinfo {year} {2015})}\BibitemShut {NoStop}%
\bibitem [{\citenamefont {Schwabl}(2006)}]{schwabl}%
  \BibitemOpen
  \bibfield  {author} {\bibinfo {author} {\bibfnamefont {F.}~\bibnamefont
  {Schwabl}},\ }\href@noop {} {\emph {\bibinfo {title} {Statistical Mechanics,
  2nd Edition}}}\ (\bibinfo  {publisher} {Springer-Verlag, Berlin Heidelberg},\
  \bibinfo {year} {2006})\BibitemShut {NoStop}%
\bibitem [{\citenamefont {L.D.Landau}\ and\ \citenamefont
  {E.M.Lifshitz}(2013)}]{landau}%
  \BibitemOpen
  \bibfield  {author} {\bibinfo {author} {\bibnamefont {L.D.Landau}}\ and\
  \bibinfo {author} {\bibnamefont {E.M.Lifshitz}},\ }\href@noop {} {\emph
  {\bibinfo {title} {Statistical Physics, 3rd edition Part 1}}}\ (\bibinfo
  {publisher} {Butterworth-Heinemann},\ \bibinfo {year} {2013})\BibitemShut
  {NoStop}%
\bibitem [{\citenamefont {Walter~Greiner}\ and\ \citenamefont
  {St{\"o}cker}(1995)}]{walter}%
  \BibitemOpen
  \bibfield  {author} {\bibinfo {author} {\bibfnamefont {L.~N.}\ \bibnamefont
  {Walter~Greiner}}\ and\ \bibinfo {author} {\bibfnamefont {H.}~\bibnamefont
  {St{\"o}cker}},\ }\href@noop {} {\emph {\bibinfo {title} {Thermodynamics and
  Statistical Mechanics}}}\ (\bibinfo  {publisher} {Springer-Verlag New York},\
  \bibinfo {year} {1995})\BibitemShut {NoStop}%
\bibitem [{\citenamefont {Huang}(1987)}]{huang}%
  \BibitemOpen
  \bibfield  {author} {\bibinfo {author} {\bibfnamefont {K.}~\bibnamefont
  {Huang}},\ }\href@noop {} {\emph {\bibinfo {title} {Statistical Physics, 2nd
  edition}}}\ (\bibinfo  {publisher} {Wiley},\ \bibinfo {year}
  {1987})\BibitemShut {NoStop}%
\bibitem [{\citenamefont {Harris}(1975)}]{harris}%
  \BibitemOpen
  \bibfield  {author} {\bibinfo {author} {\bibfnamefont {E.}~\bibnamefont
  {Harris}},\ }\href@noop {} {\emph {\bibinfo {title} {Introduction to Modern
  Theoretical Physics, Vol. 2}}}\ (\bibinfo  {publisher} {John Wiley},\
  \bibinfo {year} {1975})\BibitemShut {NoStop}%
\end{thebibliography}%

\end{document}